\documentclass[aps,prl,reprint,amsmath,amssymb]{revtex4-2}

\usepackage{graphicx}
\usepackage{booktabs}
\usepackage{siunitx}
\usepackage{hyperref}

\DeclareSIUnit{\ton}{ton}
\DeclareSIUnit{\MUSD}{MUSD}

\begin{document}

\title{Will AstroForge Collapse the PGM Market?}

\author{Robert T. Nachtrieb}
\affiliation{Senior Lecturer in System Dynamics, MIT Sloan School of Management}
\author{Steven J. Smith}
\affiliation{VP MIT Alumni System Dynamics}
\date{2026-06-24}

\begin{abstract}
AstroForge seeks to mine platinum group metals (PGM) from asteroids.
Asteroid reserves appear to be unlimited, and at current market price the
gross margin of asteroid mining would be very high.  It is natural to ask:
when AstroForge successfully demonstrates economic space mining of PGM,
will they cause the PGM market to collapse?  We answer the question with a
non-steady system dynamics model of the PGM market.  We find that the market
price for PGM will eventually drop towards the much lower cost of asteroid
mining, but only after the entire supply has shifted off-world.  In the
meanwhile, huge fortunes will be made.  And everybody on Earth will benefit
from new applications of lower-price PGM.
\end{abstract}

\maketitle

\section{Introduction}

Platinum Group Metals (PGMs) serve as the backbone for both current
industrial standards and the burgeoning green energy transition.
Traditionally, the automotive sector has been the primary consumer of these
materials, utilizing approximately 90\% of global PGM supply in the
production of catalytic converters to mitigate harmful exhaust
emissions~\cite{USGS20}.  However, as the global economy pivots from
internal combustion engines to electric and hydrogen-based platforms, the
demand profile is shifting.  The hydrogen energy sector, which relies on PGMs
for the production and utilization of hydrogen fuel, is projected to become
the largest end-market for platinum by 2040~\cite{Fla24}.  Beyond these
sectors, PGMs remain indispensable in electronics, chemical processing, and
high-end jewelry.

The PGM ``basket'' is dominated by palladium and platinum; annual production
is summarized in Table~\ref{tab:pgm_production}.  Production remains
concentrated: in 2023 and 2024, global output reached 208 and 190 metric
tons of palladium, and 179 and 170 metric tons of platinum, respectively.
Total terrestrial reserves are estimated at roughly 100{,}000 metric tons,
with the vast majority localized in Russia and South Africa's Bushveld
Complex~\cite{USGS20}.  Despite these reserves, the industry faces a
structural deficit.  Demand consistently outstrips supply~\cite{Zha24} and
is forecast to grow at a Compound Annual Growth Rate (CAGR) of 4.72\%
through 2031, driven largely by the green energy mandate~\cite{Mor26}.

\begin{table*}[htbp]
\caption{\label{tab:pgm_production}%
  Summary of PGM mine production (kg/y) and reserves (kg).
  Source:~\cite{USGS20,Sur}.}
\begin{ruledtabular}
\begin{tabular}{lrrrrr}
  & \multicolumn{4}{c}{Mine production} & \\
  \cmidrule(lr){2-5}
  & \multicolumn{2}{c}{Palladium} & \multicolumn{2}{c}{Platinum}
  & PGM reserves \\
  \cmidrule(lr){2-3}\cmidrule(lr){4-5}
  Country & 2023 & 2024$^{e}$ & 2023 & 2024$^{e}$ & \\
  \hline
  United States   &  10{,}300 &  8{,}000 &   3{,}040 &   2{,}000 &     820{,}000 \\
  Canada          &  16{,}100 & 15{,}000 &   5{,}170 &   5{,}200 &     310{,}000 \\
  Russia$^{e}$    &  87{,}000 & 75{,}000 &  21{,}000 &  18{,}000 &  16{,}000{,}000 \\
  South Africa    &  74{,}900 & 72{,}000 & 125{,}000 & 120{,}000 &  63{,}000{,}000 \\
  Zimbabwe        &  15{,}900 & 15{,}000 &  19{,}200 &  19{,}000 &   1{,}200{,}000 \\
  Other countries &   4{,}200 &  4{,}200 &   5{,}710 &   4{,}600 & N/A \\
  \hline
  World total     & 208{,}000 & 190{,}000 & 179{,}000 & 170{,}000 & $>$81{,}000{,}000 \\
\end{tabular}
\end{ruledtabular}
{\footnotesize \textbf{World resources:} PGMs are estimated to total more than
100 million kilograms.  The largest resources and reserves are in the
Bushveld Complex in South Africa.}
\end{table*}

This supply–demand imbalance has kept PGM prices high, yet terrestrial
mining operations are struggling to remain profitable.  Commodity market
volatility has compressed average gross margins to single digits, with an
estimated 25\% of producers operating at negative margins below certain spot
prices~\cite{Ste23}.  These economic pressures are exacerbated by physical
constraints: as existing mine shafts reach greater depths, the costs
associated with refrigeration, reefing, and power requirements climb
exponentially~\cite{Jac22,Mor26}.  We are approaching a geological ``event
horizon'' where the energy and capital required for extraction may soon
exceed the market value of the metals themselves.

Asteroid mining presents a transformative solution to these terrestrial
limitations, offering superior ore grades and virtually limitless volume.
While terrestrial ``high-grade'' ores typically yield 4 to 10 parts per
million (ppm), analysis of iron meteorites suggests metallic asteroids
contain PGM concentrations ranging from 100~ppm to as high as
187~ppm~\cite{Abb18}.  The scale of this resource is difficult to overstate:
a single metallic asteroid approximately one kilometer in diameter could
contain more PGMs than have been harvested in all of human history, plus all
known remaining Earth-based reserves~\cite{Ros01}.  The valuation of such a
single body reaches into the trillions of dollars~\cite{Yar22}.

Beyond economics, moving extraction off-planet offers a significant
environmental and social dividend.  Terrestrial mining is synonymous with
ecological degradation, including acid mine drainage and the contamination of
water tables with arsenic and lead~\cite{Yar22}.  Asteroid mining eliminates
these local environmental impacts and bypasses the ``not in my backyard''
(NIMBY) opposition and land-use conflicts associated with tailings
dams~\cite{Son01}.  Furthermore, the transition to highly automated
space-based mining could reduce global reliance on dangerous artisanal mining
practices often linked to human rights abuses in developing
nations~\cite{Yar22}.

Leading this transition is AstroForge, a commercial pioneer aiming to
decouple PGM supply from Earth's terrestrial constraints.  By targeting
metallic asteroids with projected profit margins of roughly 85\%---a stark
contrast to the 7\% margins seen in terrestrial mines---AstroForge is
attempting to prove the economic case for deep space resources~\cite{Ast26}.
Their strategy favors high-frequency, low-cost iteration; the company's
initial spacecraft were developed for approximately \$3.5 million, a fraction
of the \$95 million typically required by government space agencies.
Following the 2025 launch of their first deep space vehicle, \textit{Odin},
the company is currently preparing its 2026 DeepSpace-2 mission.  This
mission aims to execute the first private landing on a celestial body beyond
Earth's gravity well to provide definitive ground-truth validation of PGM
concentrations~\cite{Sno25}.

The natural question is: if AstroForge---and presumably other follow-on
ventures---successfully demonstrates that asteroid mining of PGM is viable,
will the PGM market inevitably collapse?  In a recent podcast, AstroForge
addressed this very topic.  The answer given was ``no'', but the rationale
was weak.  This is understandable, since most economic
textbooks~\cite{PR09} treat supply and demand curves with equilibrium
arguments, qualified by ``short term'' and ``long term.''  But the transition
dynamics are exactly what we are interested in.  In this note we attempt to
answer the question by developing a non-steady system dynamics~\cite{Ste00}
model of the PGM market.  We expect that ``yes'', the PGM market will
collapse.  Total market profit will increase by about a factor of eight
before collapsing to less than half the initial value.  The asteroid miners
will be the winners in the profit run-up, and the terrestrial miners will be
the losers.  And to the extent that PGM enables new technologies
(electronics, electric cars, \textit{etc.}), we will all be the winners.

\section{The Business of Asteroid Mining}

AstroForge's operational roadmap for mining Platinum Group Metals from
asteroids follows a sequential process of target validation, magnetic
anchoring, photonic extraction, and refined material return.  This
architecture is designed to avoid the prohibitively high costs of returning
waste rock to Earth or the technical complexities of mechanical drilling in
microgravity.

The process begins with the identification of Near-Earth Asteroids (NEAs)
exhibiting metallic spectral signatures.  Because ground-based observations
cannot definitively confirm the presence of high-concentration ore bodies,
AstroForge's plan utilizes precursor missions to validate targets.  The
company's roadmap involves sending scout spacecraft---such as the \textit{Odin}
and \textit{DeepSpace-2} vehicles---to rendezvous with candidate asteroids.
\textit{DeepSpace-2} is specifically designed to attempt the first commercial
landing on a body outside Earth's gravity well to verify the surface
composition and ensure the target is indeed a PGM-rich metallic body rather
than a ``rubble pile'' or non-metallic rock~\cite{Ast26,Sno25,Epi42}.  The
mission architecture utilizes high-energy launch trajectories, such as
Trans-Lunar Injection (TLI), and leverages lunar gravity assists to slingshot
the vehicle toward the target asteroid in deep space~\cite{Epi42}.

Upon arrival and matching rotation with the asteroid, the mining vehicle
faces the challenge of securing itself in a microgravity environment where
standard drilling forces would push the spacecraft away from the surface.
AstroForge addresses this by targeting ``M-type'' (metallic) asteroids, which
are composed primarily of iron and nickel, and using spacecraft equipped with
landing legs featuring switchable ``clamping'' magnets.  These magnets allow
the vehicle to adhere firmly to the iron-rich surface of the asteroid,
providing a stable platform for extraction operations without complex
mechanical gripping systems~\cite{Epi49,Epi45}.

Once anchored, the extraction system utilizes high-power lasers to vaporize
and eject regolith from the asteroid's surface into a collection system with
minimal physical resistance.  Returning raw asteroid material to Earth is
economically unfeasible due to the low density of PGMs within the bulk
iron-nickel matrix.  Therefore, the AstroForge model relies on in-situ
refining---separating the valuable PGMs from the lower-value iron and nickel
waste slag in deep space.  This process concentrates the ore, aiming to
produce a material (similar to a ``dor\'{e} bar'') with a purity of
approximately 60\% or higher~\cite{Epi45}.

The final phase involves returning only the concentrated high-value materials
to Earth.  The mining vehicle essentially functions as a return capsule; the
refined PGMs are loaded into the body of the spacecraft, which is equipped
with a standard heat shield.  The vehicle flies back to Earth and performs an
atmospheric reentry, protecting the payload until it reaches the ground for
recovery and final sale on the terrestrial market~\cite{Epi45}.

\section{Modeling the Market Transition to Asteroid PGMs}

Once AstroForge demonstrates that it is possible to recover PGM from
asteroids at eighty percent gross margin, profit motive will inspire other
companies to do the same.  Increased supply will indeed start to reduce
prices, but as long as some terrestrial sources remain the market price will
stay high enough that asteroid mining will enjoy healthy margins.  Asteroid
reserves are practically unlimited, so investment can continue until
essentially all terrestrial demand for PGM is supplied off-world.  Eventually
commodity dynamics will drive the gross margins once again down to equilibrium
values, around ten percent.

A PGM market dynamic simulation can be created to get a sense of the
trajectory from terrestrial to all off-world mining.  Three representative
points in the trajectory are considered to bound the relevant parameter
ranges:

\begin{enumerate}
  \item \textbf{Initial conditions}, representative of the PGM market today,
    when all mining is terrestrial.  Market efficiencies have driven the gross
    margin down to $\mathrm{gm} = 0.1$.

  \item \textbf{Peak conditions}, during the transition between terrestrial
    and off-world mining when gross margin peaks.  The costs are reflective of
    off-world mining, yet the market price has not yet fallen.  The average
    margin is reflective of the new technology: $\mathrm{gm} = 0.8$.

  \item \textbf{Final conditions}, when all mining has moved off-world.  Once
    again market efficiencies have driven the gross margin down to
    $\mathrm{gm} = 0.1$.
\end{enumerate}

For simplicity, it is assumed the market demand remains constant at all three
points.  The relevant market variables and their symbols are collected in
Table~\ref{tab:variables}, and the governing relations are

\begin{align}
  R   &= Q \cdot P ,                      \label{eq:revenue}\\
  E   &= Q \cdot C ,                      \label{eq:expenses}\\
  \Pi &= R - E ,                          \label{eq:profit}\\
  \mathrm{gm} &= \frac{\Pi}{R} = 1 - \frac{C}{P} , \label{eq:gm}\\
  P   &= \frac{C}{1-\mathrm{gm}} .       \label{eq:price}
\end{align}

\begin{table}[htbp]
\caption{\label{tab:variables}Market variable definitions.}
\begin{ruledtabular}
\begin{tabular}{lll}
  Description      & Symbol          & Units   \\
  \hline
  Market volume    & $Q$             & ton/y   \\
  Market price     & $P$             & M\$/ton \\
  Cost             & $C$             & M\$/ton \\
  Market revenue   & $R$             & M\$/y   \\
  Market expenses  & $E$             & M\$/y   \\
  Market profit    & $\Pi$           & M\$/y   \\
  Gross margin     & $\mathrm{gm}$   & ---     \\
\end{tabular}
\end{ruledtabular}
\end{table}

Table~\ref{tab:envelope} presents a ``back of the envelope'' sketch of the
trajectory of market dynamics.  Market profit peaks at eight times the
initial value before falling to less than one quarter the original value.

\begin{table*}[htbp]
\caption{\label{tab:envelope}%
  Back-of-the-envelope sketch of trajectory of market dynamics.}
\begin{ruledtabular}
\begin{tabular}{llllll}
  Description           & Symbol & Units    & Initial & Peak   & Final \\
  \hline
  Market demand         & $Q$    & ton/y    & 400     & 400    & 400   \\
  Market price          & $P$    & MUSD/ton & 60      & 60     & 11    \\
  Average gross margin  & GM     & ---      & 0.1     & 0.8    & 0.1   \\
  Average cost          & $C$    & MUSD/ton & 54      & 12     & 10    \\
  Market revenue        & $R$    & MUSD/y   & 24{,}000& 24{,}000& 4{,}444\\
  Market profit         & $\Pi$  & MUSD/y   & 2{,}400 & 19{,}200& 444  \\
  Steady?               &        &          & Yes     & No     & Yes   \\
\end{tabular}
\end{ruledtabular}
\end{table*}

Expanding the model into a dynamic simulation using System Dynamics allows us
to capture the market transitions from terrestrial to off-world mining,
depicted in Table~\ref{tab:envelope}, but with higher fidelity.

\subsection{Simulating PGM Supply Capacity}

\begin{figure}[htbp]
  \includegraphics[width=\columnwidth]{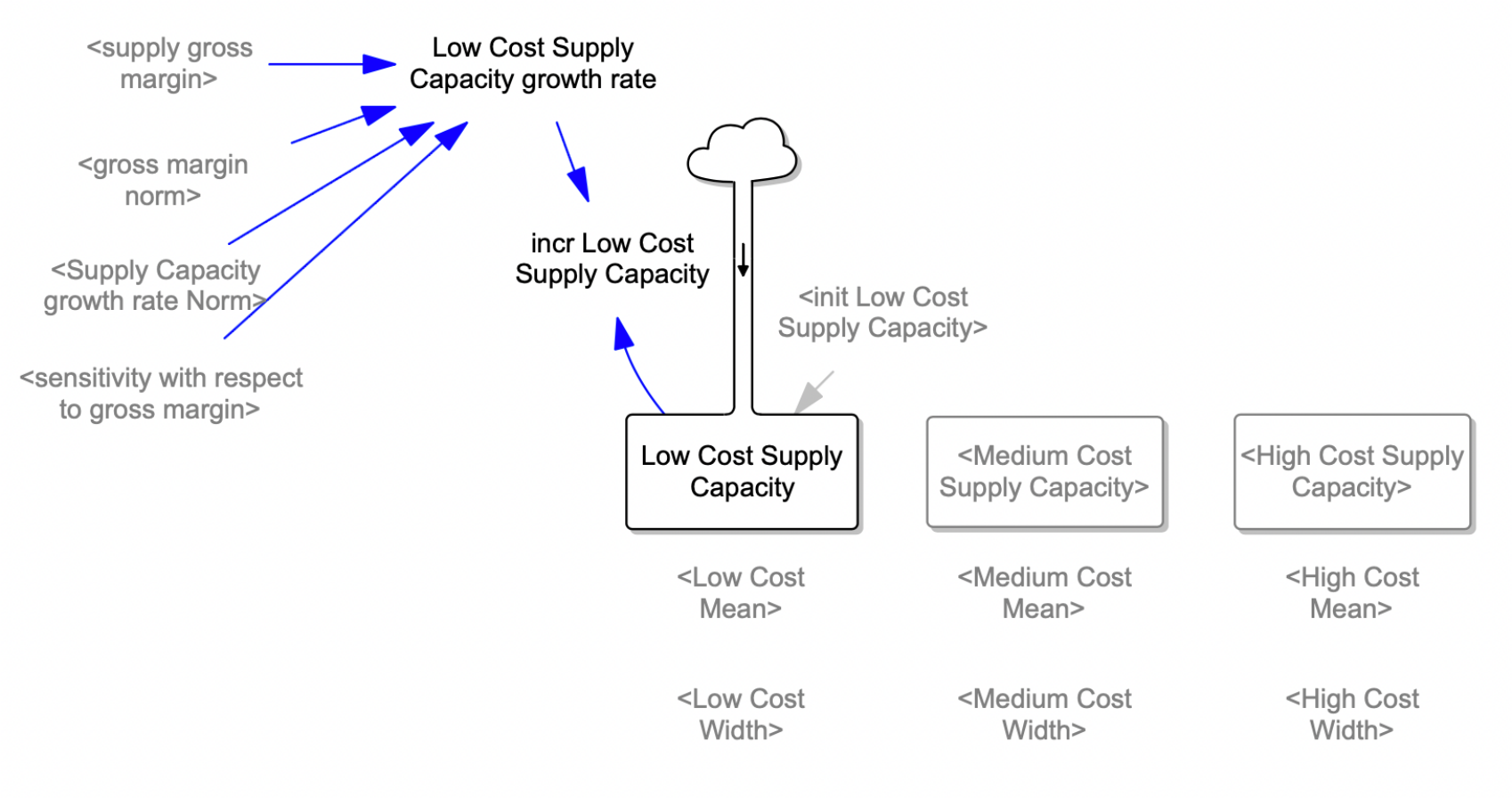}
  \caption{\label{fig:stock_flow}%
    Model of the Low Cost Supply Capacity of PGMs from asteroid mining.}
\end{figure}

Figure~\ref{fig:stock_flow} presents a stock and flow model of the PGM
supply, categorized by cost as low, medium, or high.  The bulk of the
terrestrial supply---in South Africa and Russia in today's market---is
assumed to be a Medium Cost Supply (MC).  The MC stock capacity, median cost,
and standard deviation of cost are all assumed fixed.  The High Cost Supply
(HC) is characteristic of the PGM supply found elsewhere terrestrially, such
as in the United States.  HC capacity is much smaller than MC, and the median
costs are higher, but also fixed.  In this model, the off-world asteroid
mining is considered Low Cost Supply (LC) and is variable based on the flow
generated by asteroid mining entities such as AstroForge.  LC capacity is a
new supply stock which starts small and grows through investment attracted by
high gross margins relative to an industry norm, as represented.  The LC
median cost and standard deviation of cost are assumed fixed.

The market's natural push to maximize the gross margin [Eq.~(\ref{eq:gm})]
will continuously push cost $C$ lower.  The current terrestrial-only market
favors the Medium Cost supply in Russia and South Africa over the High Cost
supply elsewhere.  Likewise, as the Low Cost supply capacity increases due to
increased asteroid mining, the favored supply will move from MC to LC.

\begin{figure}[htbp]
  \includegraphics[width=\columnwidth]{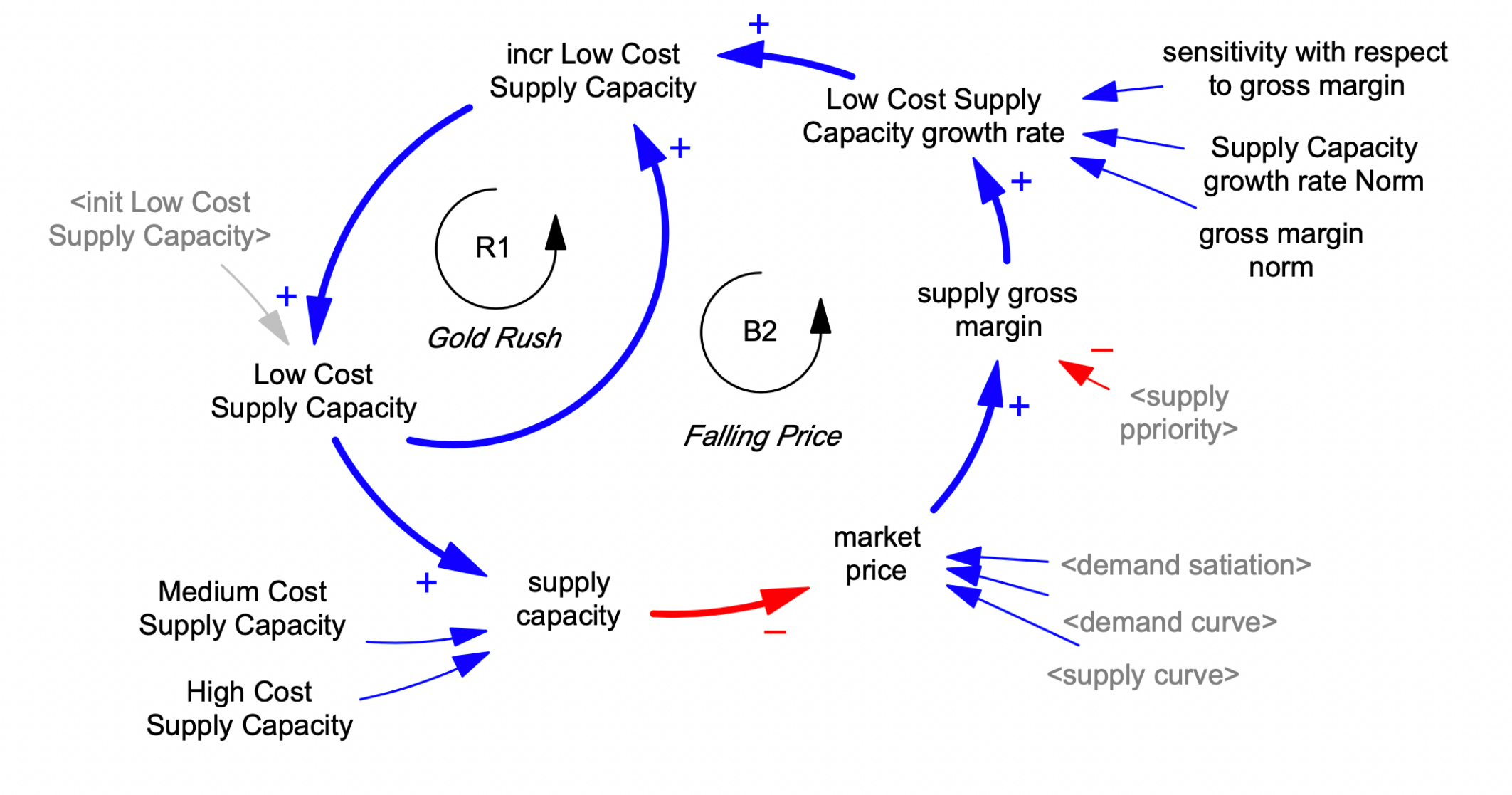}
  \caption{\label{fig:cld}%
    Asteroid PGM mining supply causal loop diagram.}
\end{figure}

The structure of the dynamics of the LC capacity can be described with a
Causal Loop Diagram, as presented in Figure~\ref{fig:cld}.

AstroForge's demonstration of low-cost, unlimited mining of PGMs kicks off
reinforcing feedback loop R1 \textit{Gold Rush}: the initial gross margin is
much higher than the industry average, which attracts exponentially growing
investment.  Eventually the low cost supply capacity is large enough to start
to depress the market price, which triggers a balancing feedback loop B2
\textit{Falling Price}, which reduces the growth rate of investment.  The
attractiveness of investment is proportional to the spread between the gross
margin of the low cost capacity and the industry norm.

\subsection{Solving for PGM Market Price}

To solve for the market price of PGMs as the supply shifts from terrestrial
to off-world mining, we use the built-in Vensim function
\texttt{FIND\_MARKET\_PRICE} to satisfy the allocation of supply to demand.
This is a non-trivial---but solved---problem.  To quote from
Ventana~\cite{VAll}, a good allocation formulation must satisfy six
properties:

\begin{enumerate}
  \item \textbf{Conservation of Matter.}  The amount received by the
    demanders (summed across all demanders) must equal the amount provided by
    the suppliers (summed across all suppliers).

  \item \textbf{Nonnegative.}  All quantities allocated must be positive or
    zero.

  \item \textbf{Conservation of Intent.}  No supplier shall provide more than
    it desired to supply and no demander shall receive more than it has chosen
    to demand. (Less, however, is possible.)

  \item \textbf{No Loopholes.}  Under conditions of adequate supply, each
    demander should receive its stated demand.  Under conditions of adequate
    demand, each supplier should supply its stated supply.

  \item \textbf{Clear Differentiation.}  When there is insufficient supply,
    extremely low-priority demanders should receive little or nothing and
    extremely high-priority suppliers should receive everything or almost
    everything.  Conversely, when there is excess supply, extremely
    unattractive suppliers should provide little or nothing and extremely
    attractive suppliers should provide the bulk of the demand.

  \item \textbf{Continuity.}  Small changes in priorities, supply, and demand
    should cause small changes in the resulting allocations.  Smoothness,
    which requires that small changes cause only small changes in the
    derivative of the allocations, is also often desirable.
\end{enumerate}

With some adjustment and manipulation, allocation logic can be made to pass
properties 1 through 4, but satisfying property 5 is very hard.  The
formulation in Vensim satisfies all six.

The many-to-many allocation problem takes the demand preferences of one or
more consumers and the supply preferences of one or more providers and
matches these~\cite{VMany}.  We discuss this as finding the price at which
the supply, summed across all providers, equals the demand, summed across all
consumers.  If one is working with true supply and demand curves, that price
is the standard market clearing price.  We first find the price that clears
the market with \texttt{FIND\_MARKET\_PRICE}, then compute the amount
supplied by the different providers using \texttt{SUPPLY\_AT\_PRICE} and the
amount demanded by the different consumers using \texttt{DEMAND\_AT\_PRICE}.

\begin{figure}[htbp]
  \includegraphics[width=\columnwidth]{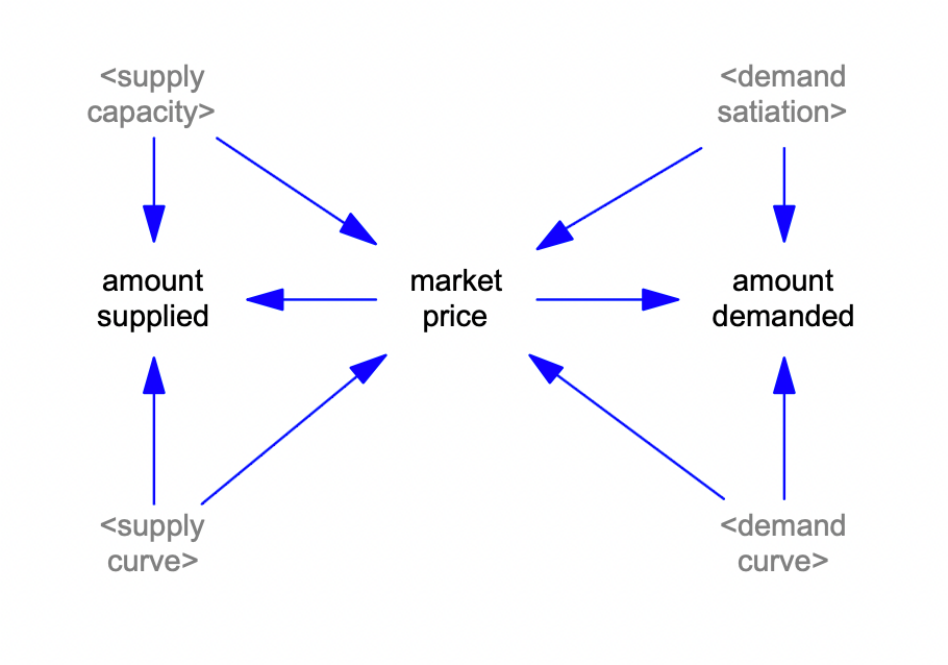}
  \caption{\label{fig:many_to_many}%
    Solving many-to-many allocation~\cite{VMany}.}
\end{figure}

\section{Results}

\begin{figure*}[htbp]
  \includegraphics[width=\textwidth]{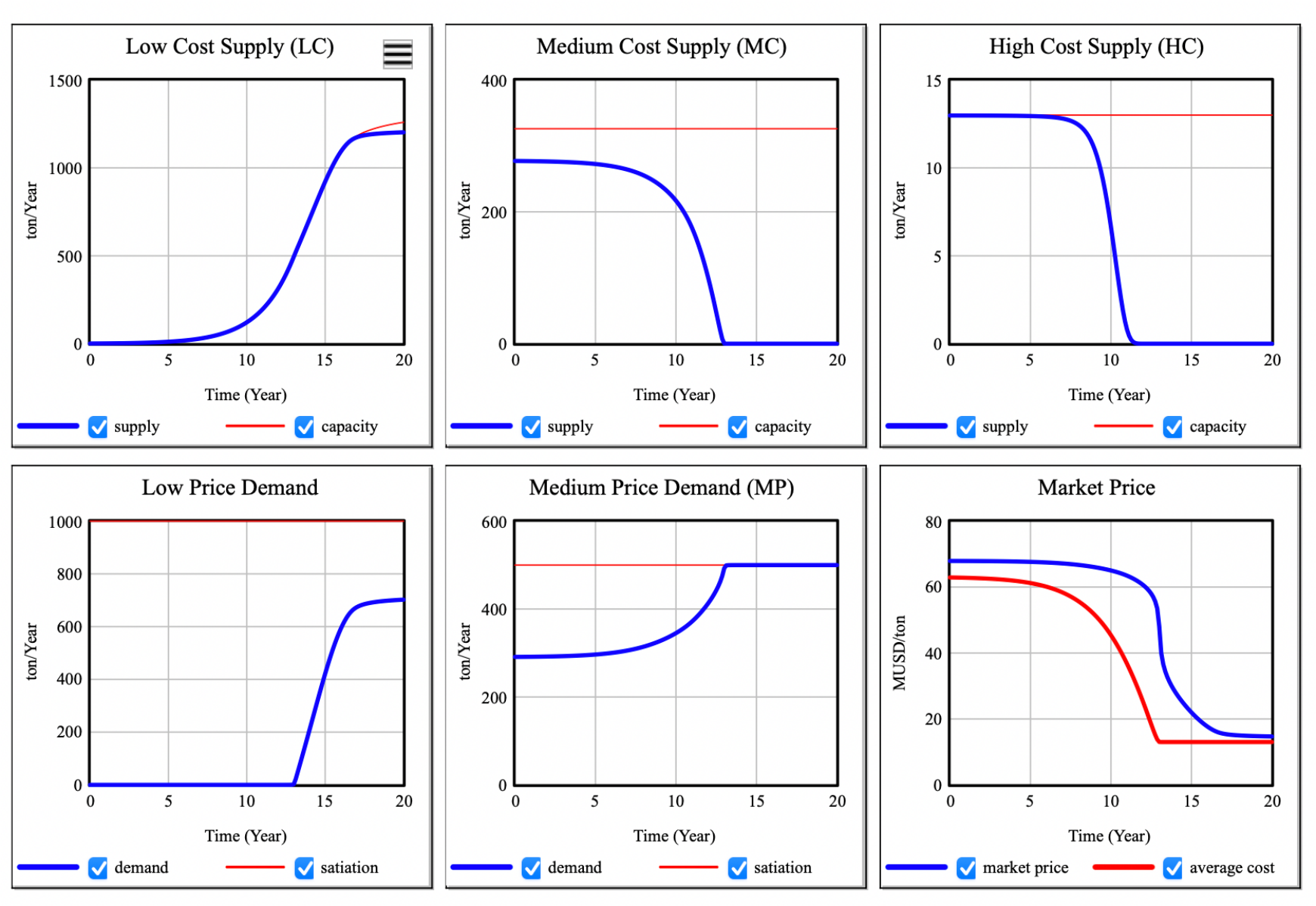}
  \caption{\label{fig:results}%
    Example scenario trajectory of PGM demand, supply, and market price as
    supply moves from terrestrial to asteroid mining.}
\end{figure*}

Using the System Dynamics model, an example scenario trajectory is presented
in Figure~\ref{fig:results}.  After the first demonstration of economic
mining of PGM on asteroids (low cost supply, LC) at $t = 0$, the appeal of
high margins attracts increasing investment resulting in exponential growth of
capacity (top left panel).  Capacity growth starts to slow when supply is
sufficient to start to depress market price (bottom right panel).  The
declining price has first impacted the thin margins of the US PGM mines
(high cost supply, HC), reducing the market share (top right panel).
Continued investment in low cost supply continues large enough to take market
share from the bulk terrestrial PGM mines (medium cost supply, MC).
Declining market price stimulates an increased demand (bottom center panel).
Eventually the market price drops down near the cost of the low cost supply,
the investment rate drops, and the low cost capacity stabilizes (top left
panel).  Growth of a new, low-price market segment (lower left panel) means
the ultimate supply quantity is much larger than the initial quantity.

\section{Discussion}

The potential for AstroForge to disrupt the Platinum Group Metals market
rests on its ability to successfully launch deep-space vehicles, intercept
metallic asteroids, and return refined materials to Earth.  If successful at
a sufficiently low cost point, the company will enter a market where prices
are currently anchored to expensive terrestrial operations, ensuring a
projected gross margin of approximately 85\%.

Our simulation model highlights that this ``Gold Rush'' phase is a product of
non-steady-state dynamics.  As AstroForge demonstrates the economic
viability of asteroid mining, the high returns will inevitably attract
competition from other commercial aerospace entities.  This influx of capital
triggers the Reinforcing Loop (R1), leading to exponential growth in Low Cost
Supply (LC) capacity.

As this supply reaches the market, it will systematically displace
terrestrial sources based on their cost structures:

\begin{itemize}
  \item \textbf{High-Cost (HC) Suppliers.}  Initial price declines will first
    impact thin-margin operations, such as those in the United States, causing
    them to lose market share early in the transition.

  \item \textbf{Medium-Cost (MC) Suppliers.}  Continued investment in
    asteroid mining will eventually take market share from the bulk of
    terrestrial production currently localized in South Africa and Russia.
\end{itemize}

Ultimately, the Balancing Loop (B2) will take effect as supply abundance
depresses market prices.  The simulation reveals that while asteroid PGM
mining is exceptionally lucrative in the short term, the market will
eventually reach a new equilibrium.  In this final state, market prices will
drop toward the actual cost of off-world extraction, and gross margins will
compress back to industry norms of roughly 10\%.

The strategic takeaway is clear: the window for outsized profits is finite.
The key to the PGM ``gold rush'' is to be an early mover and capitalize on
terrestrially-anchored prices before the market corrects for the coming era
of asteroid-sourced abundance.

\section{Conclusion}

The transition from terrestrial to space-based extraction of Platinum Group
Metals represents more than a technological milestone; it is a fundamental
shift in the global economic landscape.  As this analysis demonstrates, the
successful deployment of AstroForge's mining architecture will likely
initiate a non-steady market transition where terrestrial operations are
ill-equipped to survive.

Our system dynamics model suggests that the PGM market will undergo two
distinct phases:

\begin{enumerate}
  \item \textbf{The Gold Rush Phase.}  Early movers will capitalize on the
    vast delta between the low costs of asteroid mining and the high,
    terrestrially-anchored market prices.  During this window, market profits
    are projected to peak at eight times their initial value.

  \item \textbf{The Equilibrium Collapse.}  As asteroid-sourced supply
    eventually displaces terrestrial capacity, the market price will decouple
    from Earth-based mining costs and drop toward the lower cost of off-world
    operations.  While this ``collapses'' the traditional market price, it
    stabilizes at a new steady state where gross margins return to industry
    norms of approximately 10\%.
\end{enumerate}

The implications of this shift extend beyond the balance sheets of aerospace
firms.  The displacement of terrestrial mining offers a significant
environmental and social dividend, potentially ending the ecological
degradation and human rights concerns associated with deep-shaft and
artisanal mining.  Furthermore, the resulting abundance of low-cost PGMs
will likely catalyze a new era of industrial innovation, particularly within
the green energy and hydrogen sectors.

Ultimately, the ``collapse'' predicted by our model is not a failure of the
market, but its evolution.  While terrestrial miners face a geological and
economic ``event horizon,'' the rest of the global economy stands to benefit
from an era of virtually limitless resources.  For AstroForge and its
successors, the message is clear: the greatest rewards belong to those who
can bridge the gap between Earth and the asteroids before the market corrects
for the coming abundance.


\end{document}